# Phonon-tunable THz magnonic emission in multiferroic heterostructures

Sylvain Massabeau[1], Amr Abdelsamie[1], Florian Godel[1], Filip Miljevic[1], Noela Rezi[1], Pascale Gemeiner[2], Karim Bouzehouane[1], Thomas Buttiens[1], Sukhdeep Dhillon[3], Thomas Maroutian[1,4], Jean-Marie George[1], Henri Jaffres[1], Brahim Dkhil[2], Stephane Fusil[1], Vincent Garcia[1], Romain Lebrun[1]

[1]Laboratoire Albert Fert, CNRS, Thales, Université Paris-Saclay, 91767 Palaiseau, France
[2]Université Paris-Saclay, CentraleSupélec, CNRS, Laboratoire SPMS, 91190 Gif-sur-Yvette, France
[3]Laboratoire de Physique de l'Ecole Normale Supérieure, ENS, Université PSL, CNRS, Sorbonne Université, Université Paris Cité, F-75005 Paris, France
[4]Centre de Nanosciences et de Nanotechnologies, CNRS, Université Paris-Saclay, 91120 Palaiseau, France

**Abstract**

Collective excitations such as magnons and polar phonons provide natural access to the terahertz (THz) regime, but efficient generation and tunability remain elusive. Multiferroic $BiFeO_3$ combines both orders at room temperature, offering a unique platform for narrowband THz emission. Here, we achieve efficient sub-bandgap optical rectification of coupled phonon–polaritons near 2 THz in bare epitaxial thin films. In $Pt/BiFeO_3$ bilayers, we demonstrate that coupling the electromagnon branch with ultrafast strain waves, optically generated in Pt layers with various thicknesses, can produce tunable and narrowband emission between 0.4–0.8 THz. These results uncover the intertwined role of phonons, magnons, and magneto-acoustic dynamics in antiferromagnetic multiferroics, and establish these hybrid platforms as versatile engineered narrowband THz sources.

**Introduction**

Collective modes and quasiparticle excitation, as diverse as plasmons[1], phonons[2], low energy photons[3], magnons[4] or ferrons[5] in condensed matter all display signatures lying in the terahertz (THz) spectral range, i.e., in the 0.1-10 THz frequency window (0.4-40 meV in energy). THz emission via collective modes thus appears as a powerful probe of linear and nonlinear many-body interactions and symmetry breaking in solids, as well as on the coupling between different degrees of freedom[1,6,7]. Consequently, appropriate engineering of materials or heterostructures with low dissipation rates could lead to narrowband THz emission, providing a direct fingerprint on the ultrafast dynamics, and replace conventional THz photonic or electronic sources[8]. Combined with external stimuli such as optical, electric, magnetic, or strain, one can thus envision the subsequent excitation, manipulation and detection of these collective modes.

Antiferromagnetic materials intrinsically possess magnon modes with frequencies lying in the THz range. However, their vanishing stray field makes the manipulation and the detection of their spin dynamics[9] challenging. Theoretical proposals, as well as recent pioneering experimental studies[10,11], have proposed to use extrinsic angular momentum sources (such as spin-polarized currents[12]) or coupling with other excitations (such as magnetostriction[13–16]) to trigger their intrinsic spin dynamics. In parallel, ferroelectric materials possess polar phonons, that also lie in the THz frequency window. Modulation of the ferroelectric order is thus another control knob to tailor THz generation, for example by exciting bulk shift current at ferroelectric domain walls[17] for broadband THz emission, or by manipulating polar phonon waves for narrowband THz emission, as recently introduced with the concept of ferrons[5,18].

Room-temperature multiferroic materials such as $BiFeO_3$[19] hence gather the best of magnetic and ferroelectric material properties to investigate linear and nonlinear THz collective dynamics[20,21], as they also often exhibit strong coupling between lattice, electronic and magnetic orders[22,23]. Whilst their coherent THz spin and phonon dynamics have been largely observed in single crystals[20,24,25], approaches and mechanisms to unravel and exploit their intrinsic THz dynamics in thin films remain to be explored.

In this article, we report room-temperature narrowband THz emission in multiferroic $BiFeO_3$ via two distinct emission processes, respectively mediated by optical rectification around 2 THz in epitaxial $BiFeO_3$ thin films and by accordable ultrafast strain-waves from 0.4 to 0.8 THz in heterostructures of $Pt/BiFeO_3$. In bare thin films, we associate the presence of below-bandgap efficient optical rectification to enhanced coupled phonon-polaritons. In $Pt/BiFeO_3$ heterostructures, we evidence that the thickness of the metallic transducer enables to finely tune the frequency of the THz response with around 100 GHz resolution (limited by our setup). The pulse duration of the longitudinal acoustic phonons generated in the platinum accordingly allows to probe different regions of the electromagnon dispersion curve. Our results evidence the key role of both coherent polar phonons and magnons, and more generally of magneto-acoustic phenomena, in the ultrafast response of metal/multiferroic heterostructures.

**Results**

We first explore the THz emission properties of a 120 nm-thick film of $BiFeO_3$(111) grown on a $DyScO_3$(011) single crystal with and without a 3 nm platinum capping, using standard THz time-domain spectroscopy setup in transmission mode, while shinning 70 fs laser pulses on the sample under normal incidence, as represented in **Figure 1a**. We pump the system from the $DyScO_3$ side to avoid any THz birefringence effect from the scandate substrate[26], with a wavelength of 800 nm (1.55 eV) below the bandgap of $BiFeO_3$ (2.6-2.8 eV[27]), and record the THz signal via electro-optic detection using 200 μm and 500 μm-thick ZnTe crystals in standard configuration (detection axis ϕ = 0°). The $BiFeO_3$(111) films are epitaxially grown in a single ferroelectric domain state with a single vertical polarization variant pointing downward, as revealed by piezoresponse force microscopy (PFM) in **Fig. 1b**. In such a multiferroic $BiFeO_3$, this ferroelectric polarization is coupled to a single antiferromagnetic spin cycloid propagating in the thin film plane along the a orthorhombic axis of the substrate, as visualized by scanning NV magnetometry (**Fig. 1b**)[28]. The time-domain emission traces show clear THz oscillations for both bare $BiFeO_3$ and $Pt/BiFeO_3$, with a persistence over more than 5 ps indicating their long lifetime (**Fig. 1c**). The associated Fourier transforms (**Fig. 1d**) reveal a main peak at around 2.1 THz, and a smaller peak at around 0.7 THz only in the presence of the platinum capping. These features strikingly evidence two sources of narrowband THz emission from ultrathin multiferroic heterostructures. While the 2.1 THz peak corresponds to the reported frequency range of the THz E-phonon mode[20], the smaller 0.7 THz peak is reminiscent of the electromagnon branch[20,29] of $BiFeO_3$. Both of these two modes have only been observed in bulk crystals, and not yet in thin films of bismuth ferrite, with only one recent indirect indication of the electromagnon mode in ultrafast demagnetization experiments[11] on comparable films from the same source.

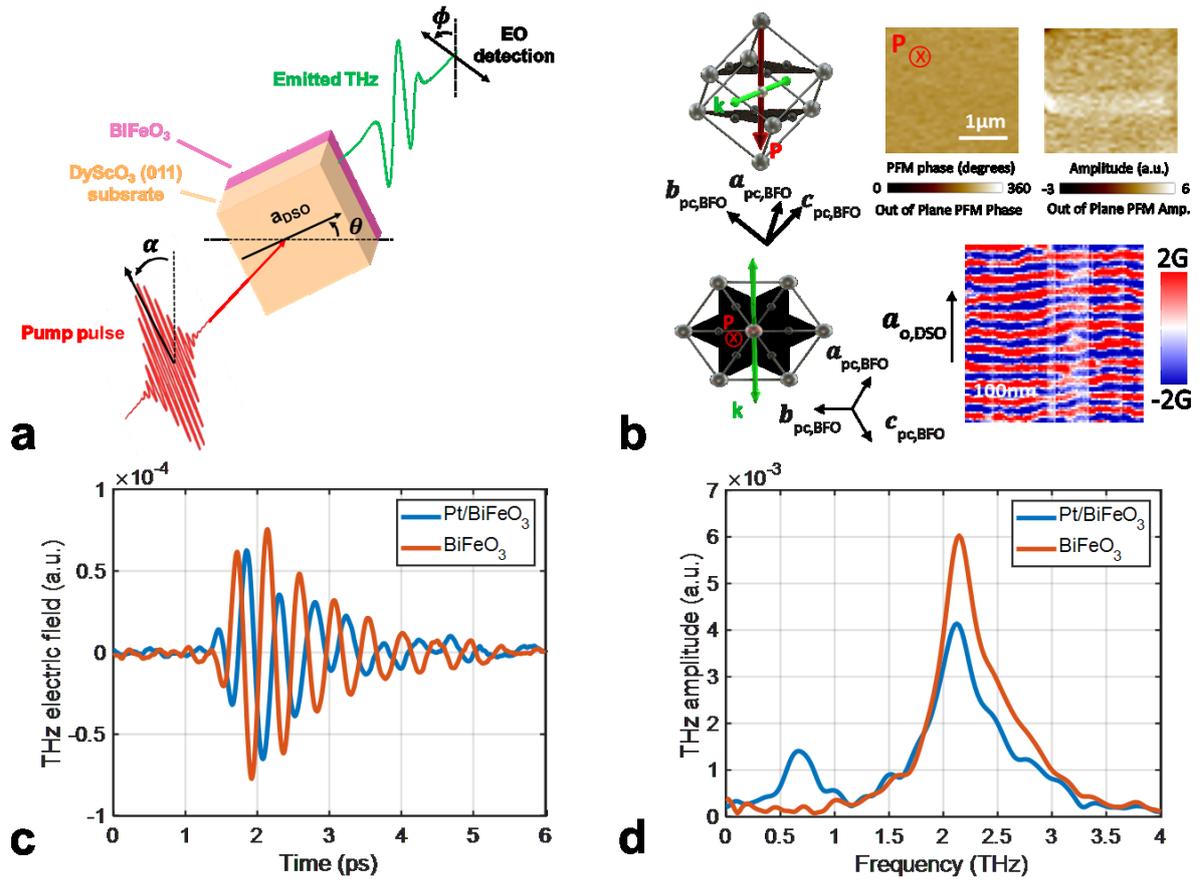

**Fig. 1 | THz emission from BiFeO$_3$ thin films and Pt/BiFeO$_3$ bilayers. a** Sketch of the experimental setup. **b** Piezoresponse force microscopy (PFM) imaging of the ferroelectric domain showing homogeneous polarization P pointing downward (top right) and NV imaging (bottom right) of the coupled single antiferromagnetic cycloid propagating along k. Sketches illustrate the BiFeO$_3$(111) pseudo-cubic unit cell in side view (top left) and its distorted hexagonal lattice in top view (bottom left). **c** Time and **d** frequency domain spectra for a 120 nm thick BiFeO$_3$(111) film grown on DyScO$_3$(011) with (blue) and without (orange) a 3 nm-thick platinum capping layer.

To understand the origins of these two contributions, we study in **Figure 2** the symmetry of the THz emission by rotating both the sample in-plane azimuthal angle θ and the incident linear laser polarization α. We decorrelate the 0.7 and 2.1 THz responses by representing the angular dependencies of the Fourier transform amplitudes at these frequencies, with and without platinum capping. We first observe that the two THz components present a uniaxial emission pattern when rotating the sample (**Fig. 2a**) and are linearly polarized (confirmed in Supplementary Information[30] by rotations of the THz detector), in line with the presence of single ferroelectric and antiferromagnetic states. Furthermore, we notice that the laser linear polarization has no sizeable impact on the low frequency contribution, whilst it strongly impacts the 2.1 THz component with a π periodicity (**Fig. 2b**). The sinusoidal dependence on the incident light polarization is consistent with both optical rectification and impulsive stimulated Raman Scattering (ISRS) processes[20,31]. Similarly to LiNbO$_3$, BiFeO$_3$ is a non-centrosymmetric material, which allows to drive polar phonons generating THz radiation by optical rectification[32]. The presence of polar phonons can enhance second-order nonlinearities[33], and thus the efficiency of the optical rectification process. However, these phonons resonances also lead to self-absorption phenomena, consequently limiting THz generation in thick films, which might explain why it has not been observed up to now in BiFeO$_3$ single crystals. Our observations highlight here that the presence of coupled phonon-polaritons provides a general approach to achieve coherent narrowband THz emission in polar materials. In contrast, the absence of polarization dependence for the 0.7 THz component indicates that it does not originate from conventional optical processes but from thermal[34] and strain stimuli, following photon absorption and hot electrons generation in the platinum layer.

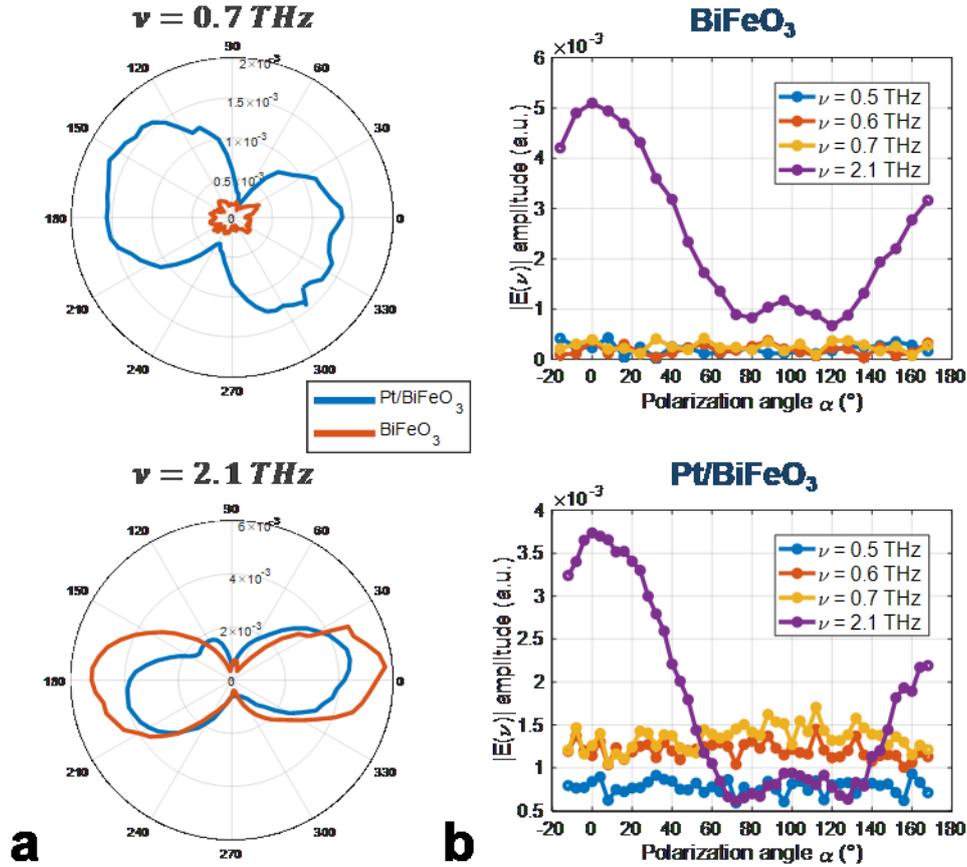

**Fig. 2 | Symmetries and angular dependencies of the THz emitted signals in bare BiFeO$_3$(111) (120 nm) and in Pt (3 nm) / BiFeO$_3$(111) (120 nm) bilayers. a** THz spectral amplitude dependencies on sample rotation θ around its surface normal axis, for the two frequency resonances, showing uniaxial emission patterns. **b** Dependence of the THz spectral amplitude at various frequencies on the linear pump polarization α, showing constant emission for the peak around 0.7 THz and a π periodicity for the signal at 2.1 THz.

Furthermore, we observe that the amplitude of the low frequency peak at 0.7 THz remains constant for BiFeO$_3$ (BFO) thicknesses ranging from 7 to 120 nm, as illustrated in **Figure 3a-b**. This feature evidences the crucial role of the metal/BiFeO$_3$ interface. On the contrary, the THz signature arising from the dynamics of the E-phonon mode around 2 THz increases in amplitude with the BiFeO$_3$ thickness. This indicates that self-absorption phenomena are still negligible up to a thickness of 120 nm, i.e., the maximum thickness allowed by epitaxial growth keeping high structural quality. The frequency response eventually does not depend on the BiFeO$_3$ thickness as it is solely defined by the material properties. One can yet notice a small redshift of the low-frequency THz component for the 50 nm-thick sample of BiFeO$_3$ capped with 5 nm of platinum, that will be discussed further later. Regarding the magnonic response at 0.7 THz, we need to assess if the peculiar spin cycloidal ordering originating from the magnetoelectric coupling is a key ingredient for the THz emission mechanism. The cycloidal ordering can be destabilized and BiFeO$_3$ can be driven to a "more standard" canted G-type antiferromagnet by epitaxial strain[23]. We therefore probe and observe similar spectra for BiFeO$_3$(001) epitaxial films grown on SmScO$_3$ substrate, in which tensile strain leads to a G-type antiferromagnetic state[35]. Raman spectroscopy confirms the presence of the E-phonon mode around 2 THz, as well as cycloid and G-type electromagnon modes in the frequency range 0.4-0.8 THz for both kinds of samples, as shown in **Fig. 3c**. The enhanced THz response for G-type BiFeO$_3$ compared with cycloidal BiFeO$_3$ is in line with the activation of other magneto-elastic coupling terms in the spin Hamiltonian of the canted phase[36].

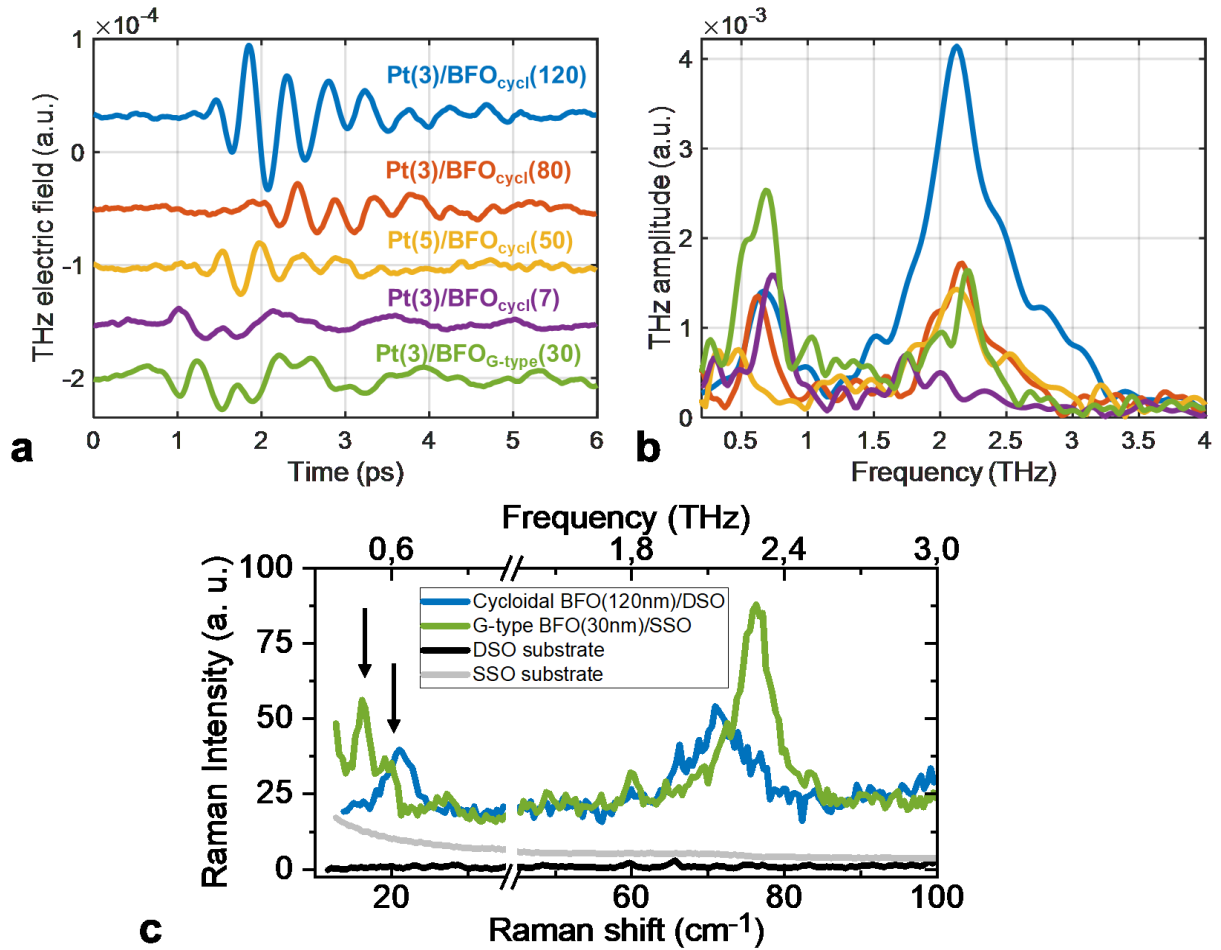

**Fig. 3 | THz modes in BiFeO$_3$ (BFO) films with different thicknesses and antiferromagnetic orders (cycloidal and G-type). a** Time and **b** frequency domain THz emission for the cycloidal antiferromagnetic order in BiFeO$_3$(111) thin films with thicknesses ranging from 7 to 120 nm, and for the 30 nm-thick G-type BiFeO$_3$(001). **c** Raman spectra of the 120 nm-thick cycloidal BiFeO$_3$, the 30 nm-thick G-type BiFeO$_3$ films, and the bare DyScO$_3$ (DSO) and SmScO$_3$ (SSO) substrates. Low frequency and high frequency spectra were acquired in crossed and parallel incident-detection polarization configurations, respectively. Arrows denote the position of the electromagnon modes.

A THz emission independent of the laser polarization, such as the 0.7 THz component, can arise from various phenomena at metal/insulator interfaces, ranging from simple optoelectronic phenomena[37,38] to more exotic magneto-acoustic processes[14,39–41]. To unravel the excitation mechanisms at play here, we investigate the role of the metallic transducer by varying the metallic material and its thickness (**Figure 4**). We switch to a thinner ZnTe crystal which has a higher signal-to-noise ratio in the low frequency range (< 2 THz) and focus on the G-type BiFeO$_3$(001) thin film displaying the highest 0.7 THz emission. As seen in **Fig. 4a-b**, we first observe a striking and surprising frequency dependency of the THz emission with the platinum thickness. This key feature is a strong indication of a magneto-acoustic generation process originating from ultrafast magnetostriction, via the strain wave generated by photon absorption and hot electron generation in the platinum. Indeed, the compressive part of the strain-wave generated in the Pt layer leaves the transducer at $d/v_{Pt}$ (d is the Pt thickness and $v_{Pt}$~4.5 nm/ps the longitudinal acoustic phonon velocity of platinum[42]) and the expansive part at $2d/v_{Pt}$, leading to the generation of a bipolar strain pulse at the metal/insulator interface[14,43]. The strain pulse entering the BiFeO$_3$ thin film consequently has a characteristic frequency defined by the inverse of the strain pulse duration. We infer a strain pulse with a duration ranging from 0.9 to 2.7 ps for Pt thicknesses varying from 2 to 6 nm, respectively, which correspond to a broadband peak in the frequency domain with a characteristic frequency ranging from 0.4 to 0.8 THz (see Supplementary Information for more details[30]). When reported on the LA phonon branch of BiFeO$_3$ (with a group velocity of 5.5 nm/ps[44]) in **Fig. 4c**, these values are in close agreement with the measured THz frequencies. Therefore, the duration of the strain pulses entering the BiFeO$_3$ thin film defines the excited part of the LA phonon spectra. However, one does not expect direct THz emission from conventional LA phonons, even in ferroelectrics, nor narrowband THz emission from a single strain pulse. Indeed, replacing for instance BiFeO$_3$ by Pb(Zr,Ti)O$_3$, a conventional non-magnetic ferroelectric material, does not give rise to any specific oscillating signature (**Fig. 4d**). In BiFeO$_3$, the THz emission is mediated by the magneto-acoustic excitation of the electromagnon branch. In line with recent theoretical predictions on collinear antiferromagnets[14], we achieve here magnon excitation via phonons when the longitudinal acoustic phonon and electromagnon branches are close. This is the case for the wavelength and frequency ranges associated with the investigated Pt thicknesses (**Fig. 4c**). In BiFeO$_3$, the electromagnon and the LA phonon branches become closer when k increases (from 0 to 500 µm$^{-1}$, i.e., down to 2 nm in wavelength). For a given pulse duration, the component with larger k vectors (lower wavelength) thus couple more efficiently with the electromagnon branch. A phenomenon of magneto-acoustic resonance then occurs and triggers narrowband THz emission.

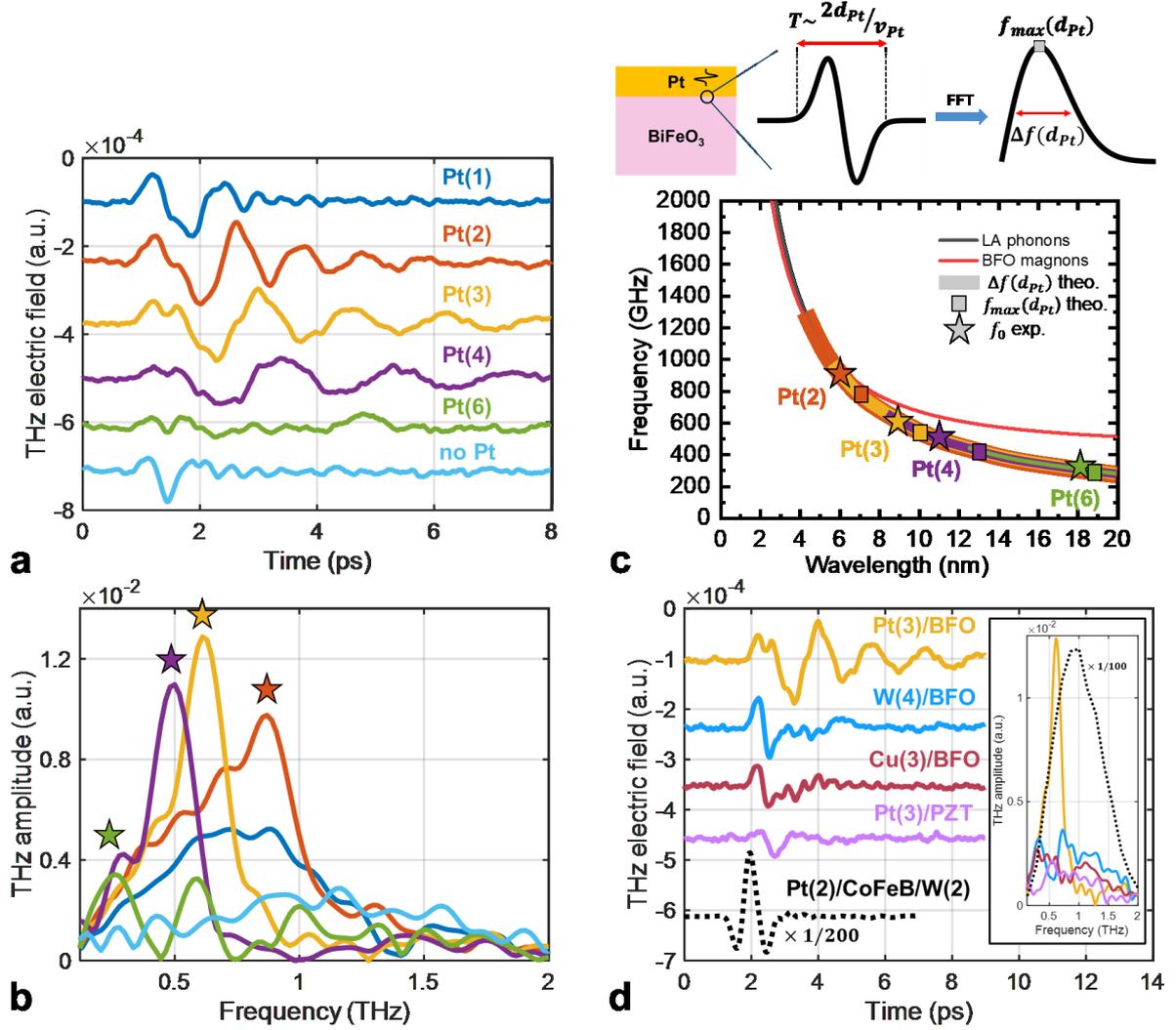

**Fig. 4 | Strain-mediated coherent THz emission in G-type Pt/BiFeO$_3$ (BFO) heterostructures. a** Time and **b** frequency domain THz emission in Pt/BiFeO$_3$ bilayers as a function of the platinum thickness, evidencing sizeable frequency dependency as denoted by colored stars at the peak maxima. **c** (Top) Sketch of the bipolar strain pulse generation in a Pt layer (with a thickness $d_{Pt}$ and a phonon group velocity of $v_{Pt}$), then injected via longitudinal acoustic (LA) phonon in the BiFeO$_3$ layer. (Bottom) Dispersion relation of the electromagnon and the LA phonon in BiFeO$_3$. Stars and squares correspond to the frequency of the THz peak extracted from (**b**) and to the theoretical maximal frequency of the phonon strain pulse, respectively. Shaded areas correspond to the accessible region of the phononic spectra for a given Pt thickness, defined by the full-width at half maximum strain pulse spectra. **d** THz emission for different metallic capping layers (W, Cu) and for another ferroelectric, Pb(Zr,Ti)O$_3$ (PZT). Platinum represents the most efficient metallic transducer compared to Cu or W. Conventional ferroelectric PZT does not show oscillating THz emission. The dashed line shows a conventional broadband metallic Pt/CoFeB/W spintronic emitter as a reference.

When capping BiFeO$_3$ with a 5 nm-thick W layer, one would expect a phase reversal of the THz time trace, in accordance with a spin-to-charge emission process, and due to the opposite spin Hall angle of W compared with Pt. Nevertheless, the low amplitude THz response does not allow to confirm with certitude such a phase reversal (**Fig. 4d**). This advocates for a THz emission linked to the electric field dynamics of the electromagnons, rather than to conventional spin-to-charge conversion. This scenario is further confirmed by adding a 2 nm-thick spin-absorbing layer of AlO$_x$ between BiFeO$_3$ and Pt (see Supplementary Information[30]), leading to the persistence of THz oscillations. Lastly, we accordingly only measure a weak THz signature for a 3 nm-thick metallic Cu capping layer, which is known to be a poor converter of infrared laser pulses into hot electrons compared with platinum[45], making it a poor strain wave generator.

Our results successfully exploit phononic and magnonic engineering to harness and control the inherent coherent THz dynamics of the prototypical antiferromagnetic multiferroic BiFeO$_3$, thereby establishing a standard approach for the development of THz magneto-acoustic devices based on magnetic materials with resonances lying in the THz and sub-THz ranges.

**Aknowledgments**

The authors thank Pascal Ruello for fruitful discussions. We acknowledge financial supports from Horizon 2020 Framework Programme of the European Commission under FET-Open Grant Agreement No. 964931 (TSAR). We are thankful for support from the French Agence Nationale de la Recherche (ANR) through the TRAPIST (ANR-21-CE24-0011), THz-MUFINS (ANR-21-CE42-0030), TATOO (ANR-21-CE09-0033), ESR/EquipEx+ program e-DIAMANT (ANR-21-ESRE-0031) and 2D-MAG (ANR-21-ESRE-0025) projects. This work is supported by a public grant overseen by the ANR as part of the "Investissements d'Avenir" program (Labex NanoSaclay; reference: ANR-10-LABX-0035). The Sesame Ile de France IMAGeSPIN project (no. EX039175) is also acknowledged. This work is also supported by a government grant managed by the ANR as part of the France 2030 investment plan from PEPR SPIN ANR-22-EXSP 0003 (TOAST) and ANR-24-EXSP-0002 (ALTEROSPIN).

**Bibliography**

1. Aupiais, I. *et al.* Ultrasmall and tunable TeraHertz surface plasmon cavities at the ultimate plasmonic limit. *Nat. Commun.* **14**, 7645 (2023).
2. Yoon, Y. *et al.* Terahertz phonon engineering with van der Waals heterostructures. *Nature* **631**, 771–776 (2024).
3. Nagatsuma, T., Ducournau, G. & Renaud, C. C. Advances in terahertz communications accelerated by photonics. *Nat. Photonics* **10**, 371–379 (2016).
4. Kampfrath, T. *et al.* Coherent terahertz control of antiferromagnetic spin waves. *Nat. Photonics* **5**, 31 (2011).
5. Choe, J. *et al.* Observation of Coherent Ferrons. Preprint at https://doi.org/10.48550/arXiv.2505.22559 (2025).
6. Mashkovich, E. A. *et al.* Terahertz light–driven coupling of antiferromagnetic spins to lattice. *Science* **374**, 1608–1611 (2021).
7. Caldwell, J. D. *et al.* Low-loss, infrared and terahertz nanophotonics using surface phonon polaritons. *Nanophotonics* **4**, 44–68 (2015).
8. Dhillon, S. S. *et al.* The 2017 terahertz science and technology roadmap. *J. Phys. Appl. Phys.* **50**, 043001 (2017).
9. Baltz, V. Antiferromagnetic spintronics. *Rev. Mod. Phys.* **90**, (2018).
10. Behovits, Y. *et al.* Terahertz Néel spin-orbit torques drive nonlinear magnon dynamics in antiferromagnetic Mn$_2$Au. *Nat. Commun.* **14**, 6038 (2023).
11. René, S. *et al.* Terahertz antiferromagnetic dynamics induced by ultrafast spin currents. *Sci. Adv.* **11**, eadx1107 (2025).
12. Cheng, R. Terahertz Antiferromagnetic Spin Hall Nano-Oscillator. *Phys. Rev. Lett.* **116**, (2016).
13. Zhuang, S., Meisenheimer, P. B., Heron, J. & Hu, J.-M. A Narrowband Spintronic Terahertz Emitter Based on Magnetoelastic Heterostructures. *ACS Appl. Mater. Interfaces* **13**, 48997–49006 (2021).
14. Azovtsev, A. V. & Pertsev, N. A. THz and sub-THz antiferromagnetic magnons via magnetoacoustic resonances excited by picosecond strain pulses in NiO. *Phys. Rev. Mater.* **8**, 044404 (2024).
15. Gabrielyan, D. A. *et al.* Ultrasonic spin pumping in the antiferromagnetic acoustic resonator α-Fe$_2$O$_3$. Preprint at https://doi.org/10.48550/arXiv.2505.22263 (2025).

16. Chen, C. *et al.* Electrical Detection of Acoustic Antiferromagnetic Resonance in Compensated Synthetic Antiferromagnets. *Phys. Rev. Lett.* **133**, 056702 (2024).
17. Guzelturk, B. *et al.* Light-Induced Currents at Domain Walls in Multiferroic $BiFeO_3$. *Nano Lett.* **20**, 145-151 (2020).
18. Tang, P., Iguchi, R., Uchida, K. & Bauer, G. E. W. Excitations of the ferroelectric order. *Phys. Rev. B* **106**, L081105 (2022).
19. Zhao, T. *et al.* Electrical control of antiferromagnetic domains in multiferroic $BiFeO_3$ films at room temperature. *Nat. Mater.* **5**, 823–829 (2006).
20. Khan, P., Kanamaru, M., Matsumoto, K., Ito, T. & Satoh, T. Ultrafast light-driven simultaneous excitation of coherent terahertz magnons and phonons in multiferroic $BiFeO_3$. *Phys. Rev. B* **101**, 134413 (2020).
21. Davydova, M. D., Zvezdin, K. A., Mukhin, A. A. & Zvezdin, A. K. Spin dynamics, antiferrodistortion and magnetoelectric interaction in multiferroics. The case of $BiFeO_3$. *Phys. Sci. Rev.* **5**, (2020).
22. Rovillain, P. *et al.* Electric-field control of spin waves at room temperature in multiferroic $BiFeO_3$. *Nat. Mater.* **9**, 975–979 (2010).
23. Haykal, A. *et al.* Antiferromagnetic textures in $BiFeO_3$ controlled by strain and electric field. *Nat. Commun.* **11**, 1704 (2020).
24. Talbayev, D., Mihály, L. & Zhou, J. Antiferromagnetic Resonance in $LaMnO_3$ at Low Temperature. *Phys. Rev. Lett.* **93**, 017202 (2004).
25. Białek, M., Magrez, A., Murk, A. & Ansermet, J.-Ph. Spin-wave resonances in bismuth orthoferrite at high temperatures. *Phys. Rev. B* **97**, 054410 (2018).
26. Yang, C.-J. *et al.* Birefringence of orthorhombic $DyScO_3$: Toward a terahertz quarter-wave plate. *Appl. Phys. Lett.* **118**, 223506 (2021).
27. Kumar, A. *et al.* Linear and nonlinear optical properties of $BiFeO_3$. *Appl. Phys. Lett.* **92**, 121915 (2008).
28. Dufour, P. *et al.* Onset of Multiferroicity in Prototypical Single-Spin Cycloid $BiFeO_3$ Thin Films. *Nano Lett.* **23**, 9073–9079 (2023).
29. De Sousa, R. & Moore, J. E. Optical coupling to spin waves in the cycloidal multiferroic $BiFeO_3$. *Phys. Rev. B* **77**, 012406 (2008).
30. Supplementary Information includes modelling of the strain transduction, fluence dependency of the two THz components, polarisation direction of the THz emitted fields, and the influence of $AlO_x$ spacer insertion between Pt and $BiFeO_3$.
31. Hlinka, J., Pokorny, J., Karimi, S. & Reaney, I. M. Angular dispersion of oblique phonon modes in $BiFeO_3$ from micro-Raman scattering. *Phys. Rev. B* **83**, 020101 (2011).
32. Carletti, L. *et al.* Nonlinear THz Generation through Optical Rectification Enhanced by Phonon–Polaritons in Lithium Niobate Thin Films. *ACS Photonics* **10**, 3419–3425 (2023).
33. Lu, Y. *et al.* Giant enhancement of THz-frequency optical nonlinearity by phonon polariton in ionic crystals. *Nat. Commun.* **12**, 3183 (2021).
34. Kholid, F. N. *et al.* The importance of the interface for picosecond spin pumping in antiferromagnet-heavy metal heterostructures. *Nat. Commun.* **14**, 538 (2023).
35. Chaudron, A. *et al.* Electric-field-induced multiferroic topological solitons. *Nat. Mater.* **23**, 905–911 (2024).
36. Rõõm, T. *et al.* Magnetoelastic distortion of multiferroic $BiFeO_3$ in the canted antiferromagnetic state. *Phys. Rev. B* **102**, 214410 (2020).
37. Metzger, T. W. J. *et al.* Separating Terahertz Spin and Charge Contributions from Ultrathin Antiferromagnetic Heterostructures. *Phys. Rev. Lett.* **135**, 076702 (2025).

38.     Yang, D. & Tonouchi, M. Understanding terahertz emission properties from a metal–insulator–semiconductor structure upon femtosecond laser illumination. *J. Appl. Phys.* **130**, 055701 (2021).

39.     Rongione, E. *et al.* Emission of coherent THz magnons in an antiferromagnetic insulator triggered by ultrafast spin–phonon interactions. *Nat. Commun.* **14**, 1818 (2023).

40.     Zhuang, S. *et al.* Hybrid magnon-phonon cavity for large-amplitude terahertz spin-wave excitation. *Phys. Rev. Appl.* **21**, 044009 (2024).

41.     Zeuschner, S. P. *et al.* Standing spin wave excitation in Bi : YIG films via temperature-induced anisotropy changes and magneto-elastic coupling. *Phys. Rev. B* **106**, 134401 (2022).

42.     Dwight E. Gray. *American Institute of Physics Handbook*. (Internet Archive, 1957).

43.     Azovtsev, A. V. & Pertsev, N. A. Excitation of high-frequency magnon modes in magnetoelastic films by short strain pulses. *Phys. Rev. Mater.* **4**, 064418 (2020).

44.     Ruello, P. *et al.* Photoexcitation of gigahertz longitudinal and shear acoustic waves in $BiFeO_3$ multiferroic single crystal. *Appl. Phys. Lett.* **100**, 212906 (2012).

45.     Bergeard, N. *et al.* Tailoring femtosecond hot-electron pulses for ultrafast spin manipulation. *Appl. Phys. Lett.* **117**, 222408 (2020).